\documentclass[10pt,conference]{IEEEtran}
\usepackage{graphics} 
\usepackage[pdftex]{graphicx}
\usepackage{times} 
\usepackage [cmex10]{amsmath} 
\usepackage{amssymb}  
\IEEEoverridecommandlockouts
\title{Information Capacity of Energy Harvesting\\ Sensor Nodes
\thanks{ R Rajesh is  with Center for Airborne Systems, DRDO, Bangalore. Vinod Sharma is with ECE dept. Indian Institute of Science, Bangalore. Pramod Viswanath is with  Electrical and Computer Engineering dept. at University of Illinois, Urbana-Champaign. This work was done when Prof. Viswanath was visiting Indian Institute of Science.  Email:rajesh81r@gmail.com, vinod@ece.iisc.ernet.in, pramodv@uiuc.edu.}
\thanks{This work is partially supported by a grant from ANRC to Prof. Sharma.}
\thanks{The visit of Prof. Viswanath is supported by DRDO-IISc Programme on Advanced Mathematical Engineering.}}
\author{R Rajesh, Vinod Sharma and Pramod Viswanath}
\begin{document}
\maketitle
\thispagestyle{empty}
\pagestyle{empty}
\begin{abstract}
Sensor nodes with energy harvesting sources are gaining popularity due to their ability to improve the network life time and  are becoming a preferred choice supporting `green communication'. We study  such a sensor node with an energy harvesting source and  compare various architectures by  which the harvested energy is used.  We find its Shannon capacity when it is transmitting its observations over an AWGN channel and  show that the capacity achieving energy management policies are related to the throughput optimal policies. We  also obtain the capacity when energy conserving  sleep-wake modes are supported and an achievable rate for the system with  inefficiencies in energy  storage.
\end{abstract}

\section{Introduction}
\label{intro}
Sensor nodes are often  deployed for monitoring a random field. These nodes are characterized by  limited battery power, computational resources  and storage space. Often, once deployed the battery of these nodes are not changed because of the inaccessibility of these nodes.  Nodes could possibly use larger batteries but with increased weight, volume and cost.  Hence when the battery of a node is exhausted, it is not replaced and the node dies. When sufficient number of nodes die, the network may not be able to perform its designated task. Thus the life time of a network is an important characteristic of a sensor network (\cite{bhardwaj}) and it depends on the life time of a node.


Recently new techniques of  increasing network life time by increasing the life time of  the battery is gaining popularity. This is made possible by energy harvesting techniques (\cite{kansal}, \cite{niyato}). Energy harvester harnesses energy from environment or other energy sources ( e.g., body heat) and converts them to electrical energy. Common energy harvesting devices are solar cells, wind turbines and  piezo-electric cells, which extract energy from the environment. Among these,  harvesting solar energy through photo-voltaic effect seems to have emerged as a technology of choice for many sensor nodes (\cite{niyato}, \cite{raghunathan}). Unlike for a battery operated sensor node, now there is potentially an \textit{infinite} amount of energy available to the node. However, the source of energy and the energy harvesting device may be such that the energy cannot be generated at all times (e.g., a solar cell).  Furthermore the rate of generation of energy can be limited. Thus one may want to match the energy generation profile of the harvesting source with the energy consumption profile of the sensor node. If the energy can be \textit{stored} in the sensor node then this matching can be considerably simplified. But the energy storage device may have limited capacity. The energy consumption policy  should be designed in such a way that the node can perform satisfactorily for a long time, i.e., energy starvation at least, should  not be the reason for the node to die. In \cite{kansal} such an energy/power management scheme is called  \textit{energy neutral operation}.

We study the Shannon capacity of such an energy harvesting sensor node transmitting over an Additive White Gaussian Noise (AWGN) Channel. We provide the capacity under various energy buffer conditions and show that the capacity achieving policies are related to throughput optimal policies(\cite{vinod1}). We also study generalizations of this system with  various inefficiencies in storage and different modes of operation.

In the following we survey the relevant literature. Early papers on energy harvesting in sensor networks are \cite{kansal1} and \cite{rahimi}. A practical solar energy harvesting sensor node prototype is described in \cite{jiang}. In  \cite{kansal} various deterministic  models for energy generation and energy consumption profiles are studied and provides conditions for energy neutral operation. In \cite{jaggi}  a sensor node is considered which is sensing certain interesting events. The authors study optimal sleep-wake cycles such that event detection probability is maximized.

 Energy harvesting can be  often divided into two major architectures. In {\it{Harvest-use}}(HU), the harvesting system directly powers the sensor node and when sufficient energy is not available the node is disabled. In {\it{Harvest-Store-Use}}(HSU) there is a storage component that stores the harvested energy and also powers the sensor node. The storage can be single or double staged (\cite{kansal},~\cite{jiang}).

Various throughput and delay optimal energy management policies for energy harvesting sensor nodes are provided in \cite{vinod1}.  These energy management policies in \cite{vinod1} are extended in  various directions in \cite{vinod2} and \cite{vinod3}. For example, \cite{vinod2} also provides some efficient MAC policies for energy harvesting nodes. In \cite{vinod3} optimal sleep-wake policies are obtained for such nodes. Furthermore, \cite{vinod4} considers jointly optimal routing, scheduling and power control policies for networks of energy harvesting nodes.



None of the above studies considers information-theoretic capacity of energy harvesting sensor nodes. We consider this problem in this paper. We also compute the capacity when the energy is also consumed in other activities at the node (e.g., processing, sensing etc) than transmission. Finally we provide the achievable rates when there are storage inefficiencies. We show that the throughput optimal policies provided in \cite{vinod1} are related to the  capacity achieving policies provided here. After submission of the first version to {\tt arxiv}, independently \cite{uluk} appeared on {\tt{arxiv}} which addresses this problem for a special case of the model considered in Theorem 1 of this paper. Their method of proof is entirely different. Our results can be useful in the context of green communication (\cite{green1},~\cite{green2}) when solar and/or wind energy can be used by a base station (\cite{wind}).


The paper is organized as follows. Section \ref{model} describes the system model. Section \ref{stability} provides the capacity for a single node under idealistic assumptions. We show that the capacity achieving policy is close to the throughput-optimal policy for the infinite buffer case.   Section \ref{simulation} takes into account the energy spent on sensing, computation etc. and proposes capacity achieving  sleep-wake schemes. Section \ref{opt} obtains efficient policies with inefficiencies in the energy storage system. Section \ref{conclude} concludes the paper.

\section{Model and notation}
\label{model} In this section we present our model for a single energy harvesting sensor node.

\begin{figure}[h]
\begin{center}
\includegraphics[height=1.4in, width=3.5in]{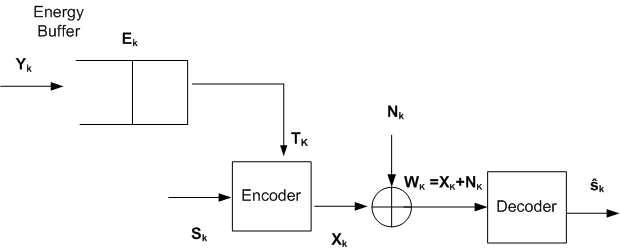}
\caption{The model} \label{fig1}
\end{center}
\end{figure}
We consider a sensor node (Fig. \ref{fig1}) which is sensing and generating data to be transmitted to a central node via a discrete time AWGN Channel.  We assume that transmission consumes most of the energy in a sensor node and ignore other causes of energy consumption (this is true for many low quality, low rate sensor nodes (\cite{raghunathan})). This assumption will be removed in Section \ref{simulation}. The sensor node is able to replenish energy by $Y_k$  at time $k$. The energy available in the node at time $k$ is $E_k$. This energy is stored in an energy buffer with an infinite capacity. 

The node uses energy $T_k$ at time $k$ which depends on $E_k$ and $T_k \le E_k$ . The process $ \{ E_k \}$ satisfies
\begin{eqnarray}
E_{k+1}  = (E_k - T_k)^{+} + Y_k. \label{eqn2}
\end{eqnarray}

We will assume that  $\{ Y_k \}$  is  stationary ergodic. This assumption is general enough to cover most of the stochastic models developed for energy harvesting.  Often the energy harvesting process will be time varying (e.g., solar cell energy harvesting will depend on the time of day). Such a process can  be approximated by piecewise stationary processes. As in \cite {vinod1}, we can indeed consider $\{Y_k\}$ to be periodic stationary ergodic.

The encoder receives a message $S$ from the node and generates an $n$-length codeword to be transmitted on the AWGN channel. The channel output  $W_k=X_k+N_k$ where $X_k$ is the channel input at time $k$ and $N_k$ is independent, identically distributed (\emph{iid}) Gaussian noise with zero mean and variance $\sigma^2$ (we denote the corresponding Gaussian density by $\mathcal{N}(0,\sigma^2))$. The decoder receives $W^n \stackrel{\Delta}{=}(W_1,...,W_n)$ and reconstructs  $S$ such that the probability of decoding error is minimized.

We will obtain the information-theoretic capacity of this channel. This of course assumes that there is always data to be sent at the sensor node. This channel is essentially different from the usually studied systems in the sense that the transmit power and coding scheme can depend on the energy available in the energy buffer at that time.

A possible generalization of our model is that the energy $E_k$ changes at a slower time scale than a channel symbol transmission time, i.e., in equation \eqref{eqn2} $k$ represents a time slot which consists of $m$ channel uses, $m \ge 1$. Due to space constraints we comment on this generalization only in Section III.





\section{Capacity for the Ideal System}
 \label{stability}
In this section we obtain the capacity of the channel with energy harvesting node under ideal conditions of  infinite energy buffer and energy consumption in transmission only.

The system starts at time $k=0$ with an empty energy buffer and $E_k$ evolves with time depending on $Y_k$ and $T_k$. Thus $\{E_k,~k \ge 0\}$ is not stationary and hence $\{T_k\}$ may also not be stationary. In this setup, a reasonable general assumption is to expect $\{T_k\}$ to be asymptotically stationary. One can further generalize it to be Asymptotically Mean Stationary (AMS), i.e.,
\begin{equation}
\lim_{n \to \infty} \frac{1}{n} \sum_{k=1}^n P[T_k \in A]= \overline{P}(A)
\end{equation}
exists for all measurable $A$. In that case $\overline{P}$ is also a probability measure and is called the \emph{stationary mean} of the AMS sequence (\cite{gray}).

If the input $\{X_k\}$ is AMS, then it can be easily shown that for the AWGN channel $\{(X_k,W_k),~k \ge 0\}$ is also AMS. Now the relevant channel capacity of our system is (\cite{gray})
\begin{equation}
\label {eqnn1}
C= \sup_{p_x} \overline{I}(X;W)= \sup_{p_x} ~ \limsup_{n \to \infty}\frac{1}{n} I(X^n,W^n)
\end{equation}
where $\{X_n\}$ is an AMS sequence, $X^n=(X_1, ..., X_n)$ and the supremum is over all possible AMS sequences $\{X_n\}$. If $R< C$ then one can find a sequence of codewords with codelength $n$ and rate $R$ such that  the average probability of error goes to zero as $n \to \infty$ (\cite{gray}).

In the following we obtain $C$ in \eqref{eqnn1} for our system. We need the following definition. Although for $C$ we need mutual information rate $\overline{I}(X;W)$, $I^{*}(X;W)$ defined below is easier to handle.

\emph{Pinsker Information Rate} (\cite{gray}): Given an AMS  random process $\{(X_n, W_n)\}$ with
 standard alphabets (Borel subsets of Polish spaces) $A_X$ and $A_W$, the Pinsker information rate is defined as
 $I^{*} (X;W)= \sup_{q,r} \overline{I} (q(X); r(W))$ where the supremum is over all quantizers $q$ of
$A_X$  and $r$ of $A_W$ and $\overline{I}(q(X);r(W)) = \limsup_{n \to \infty}  I(q(X^n);r(W^n))/n$.

It is known that, $ I^*(X;W) \le \overline{I}(X;W)$. The two are equal if the alphabets are finite.

We also need the following Lemma for proving the achievability of the capacity of the channel. This Lemma holds for $I^*$ but not for $\overline{I}$.

{\bf Lemma 1} (\cite{gray}, Lemma  6.2.2):  Let  $\{X_n, W_n\}$  be  AMS  with distribution $P$ and stationary mean $\overline{P}$. Then $I^*_P(X;W) = I^*_{\overline{P}}(X;W)$.

{\bf Theorem 1} For the energy harvesting system $C = 0.5~\log(1+\frac{E[Y]}{\sigma^2}) $.

{\bf Proof:}  \emph{Achievability}:  Let $\{X_k'\}$ be an $iid$
Gaussian sequence with mean zero and variance $E[Y]-\epsilon$ where
$\epsilon > 0$ is an arbitrarily small constant. We define the
channel input $X_k= sgn (X_k') \min( \sqrt{E_k},|X_k'|)$ where $sgn(x)=1$ if $x\ge0$ and $=-1$ if $x<0$. Then $T_k=
X_k^2 \le E_k$ and $E[T_k]=E[X_k^2]\le E[Y]-\epsilon$. Thus $E_k \to
\infty~a.s.$ and $|X_k-X_k'| \to 0~a.s$. Therefore, $\{X_k\}$ is
AMS ergodic and $\{X_k, W_k\}$ is AMS ergodic.

By using Lemma 1
 $I^*(X;W)=\sup_{q,r}~\limsup_{n \to \infty}  I(q(X^n);r(W^n))=I^*({X}',{W}')/n$
 where $I^*({X}',{W}')$ corresponds to the limiting $iid$ sequence $\{{X}_i',{W}_i'\}$
 where ${W}_i'$ is the channel output corresponding to $X_i'$.

Also, since the mutual information between two
random variables is the limit of the  mutual information between their
quantized versions (\cite{Cover04elements}),
 $I^*({X}',{W}')=I({X}',{W}')=0.5~\log(1+{(E[Y]-\epsilon)}/\sigma^2)$.
Therefore,  $I^*({X}',{W}') \le \overline{I}(X;W)= \limsup_{n \to \infty}  I(X^n;W^n)/n$ and
hence $\limsup_{n \to \infty}  I(X^n;W^n)/n= \overline{I}(X;W) \ge 0.5~\log(1+{(E[Y]-\epsilon)}/{\sigma^2})$
 for all $\epsilon>0$.

\emph {Converse  Part}:  For the system under consideration
 $\frac{1}{n} \sum_{k=1}^n T_k \le \frac{1}{n} \sum_{k=1}^n Y_k \to E[Y] ~a.s.$
Hence, if $\{X_k(s),~k=1,...,n\}$ is a codeword for message $s \in \{1,...,2^{nR}\}$ then for
 all large $n$ we must have $\frac{1}{n} \sum_{k=1}^n X_k(s)^2 \le E[Y]$ with a large probability.
Hence by the converse in the AWGN channel case,
 $\limsup_{n \to \infty} \frac{1}{n} I(X^n;W^n)\le 0.5~\log(1+{E[Y]}/{\sigma^2})$.

Combining the direct part and converse part completes the proof. ~~~~~~~~~~~~~~~~~~~~~~~~~~~~~~~~~~~~~~~~~~~~~~~~~~~~~~~~~~~~~~~~
 {\raggedleft{$\blacksquare$}}
\vspace{0.1cm}

Thus we see that the capacity of this channel is the same as that of a node with average energy constraint $E[Y]$, i.e., the hard energy constraint of $E_k$ at time $k$ does not affect its capacity.
The capacity achieving signaling in the above theorem is $iid$ Gaussian
 with zero mean and variance $E[Y]$.

If there is no energy buffer to store the harvested energy (Harvest-Use)
then $T_k=X_k^2 \le Y_k$. Thus $C=\max_{p_x}I(X;W)\le 0.5~ \log (1+ E[Y]/\sigma^2)$. The last inequality is strict unless $X_k$ is $\mathcal{N}(0,E[Y])$. Then $X^2=Y$ and hence $Y_k$ is chi-square distributed with degree 1. If $Y_k\equiv E[Y]$ then the capacity will be that of an AWGN channel with peak and average power constraint $=E[Y]$. This problem is addressed  in \cite{peak1}, \cite{peak2}, \cite{peak3} and the capacity achieving distribution is finite and discrete. Let  $X(y)$ denote a random variable  having distribution that  achieves capacity with peak power $y$. Then, for the general case the capacity of the channel is

\begin{equation}
C= E_Y[I(X(Y);W)].\label{fin}
\end{equation}
For small $y$, $X^2(y)=y$.

Thus, having energy buffer to store the harvested energy almost always strictly increases the capacity of the system under ideal conditions.

In \cite{vinod1}, a system with  a data buffer at the node which
stores data sensed by the node before transmitting it, is considered.
 The stability region (for the data buffer) for the 'no-buffer' and 'infinite-buffer'
corresponding to the harvest-use and harvest-store-use architectures are provided.
The throughput optimal policies in \cite{vinod1} are  $T_n= \min(E_n;E[Y]-\epsilon)$ for
 the  infinite energy buffer and  $T_n = Y_n$ when there is no energy buffer. Hence
 we see that the Shannon capacity achieving energy management policies provided here are
 close to the throughput optimal policies in \cite{vinod1}.  Also the capacity is the
 same as the maximum throughput obtained in the data-buffer case in \cite{vinod1} for the infinite buffer architecture.

An advantage of the above capacity/throughput optimal policies is that they are easy to implement online (atleast for the infinite buffer case).

Above we considered the cases when there is infinite
 energy buffer or when there is no buffer at all.
However, in practice often there is a finite energy buffer to store. Let $\gamma$ be the buffer size. If $E_k \le \gamma < \infty$, then the methods of Theorem 1 will allow us to upper bound the capacity of the system. Let $\{X_k'\}$ be $iid$ with capacity achieving distribution for an AWGN channel with peak power constraint $\gamma$ and average power constraint $E[Y]$. Our system has capacity close to this capacity with the signaling as $X_k=sgn(X_k')\min(\sqrt{E_k},|X_k'|)$. As $\gamma$ increases, this upper bound becomes tighter. Also, the upper bound approaches the capacity in Theorem 1. If we allow $Y_k$ to be used in slot $k$ itself (use energy before storing), then the bound can be exceeded (but will again be approached as $\gamma \to \infty)$.

Next we comment on the capacity results when \eqref{eqn2} represents $E_{k+1}$ at the end of the $k$th slot where a slot represents $m$ channel uses. In this case energy $E_k$ is available not  for one channel use but for $m$ channel uses. This relaxes our energy constraints. Thus if $E[Y]$ still denotes mean energy harvested per channel use, then for infinite buffer case the capacity remains same as in Theorem 1. However the no buffer case becomes like a finite buffer case studied above.

\section{Capacity with Processor Energy (PE)}
\label{simulation}
Till now we have assumed that all the energy that a node consumes is for transmission.
However, sensing, processing and receiving (from other nodes) also require significant energy,
 especially in recent higher-end sensor nodes (\cite{raghunathan}).
We will now include the energy consumed by sensing and processing only. We will
 see that if a node has an energy saving mode, the achievable rate can be improved. For simplicity, we will initially assume that
 the node is always in one energy mode (i.e., there is no energy saving mode).

We  assume that energy $Z_k$  is consumed by the node (if $E_k \ge Z_k$) for
 sensing and processing  at time instant $k$. For transmission at time $k$, only $E_k-Z_k$ is available.
 $ \{ Z_k \}$ is assumed a stationary ergodic sequence. The rest of the system is as in Section \ref{model}.

First we extend the achievable policy in Section \ref{stability} to
incorporate this case. The signaling scheme $\{X_k\},~
X_k=sgn(X_k')\min(\sqrt{E_k},|X_k'|)$ where $\{X_k'\}$ is $iid$
Gaussian with zero mean and variance $E[Y]-E[Z]-\epsilon$  achieves
the rate
\begin{equation}
R_{PE}= 0.5~\log\left(1+\frac{E[Y]-E[Z]-\epsilon}{\sigma^2}\right).\label{abc}
\end{equation}

If the sensor node has two modes: Sleep and Awake then the achievable rates can be improved. The sleep mode is a power saving
mode in which  the sensor only harvests energy and performs no other functions so that the energy consumption is minimal (which will be ignored). If $E_k < Z_k$ then we assume that the node will sleep at time $k$. But to optimize its transmission rate it can sleep at other times also. We consider a policy called \emph{randomized sleep policy} in \cite{vinod3}. In  this policy at every time instant $k$ with $E_k \ge Z_k$ the sensor chooses to sleep with probability $p$ independent of all other random variables.  We will see that such a policy can be capacity achieving in the present context. 

With the sleep option we will show that the capacity of this system is
\begin{equation}
C= \sup_{p_x: E[b(X)] \le E[Y]} I(X;W) \label{impe}
\end{equation}
where $b(x)$ is the cost of transmitting $x$ and equals
\begin{eqnarray*}
b(x)= \begin{cases}
x^2+\alpha,~& \text{if $|x| >0$},\\
0,~& \text{if $|x| = 0$}.
\end{cases}
\end{eqnarray*}
and $\alpha= E[Z]$. Observe that if we follow a policy that unless the node transmits, it sleeps,
 then $b$ is the cost function.  An optimal policy will have this characteristic.
  Denoting the expression in \eqref{impe} as $C(E[Y])$, we can easily check that $C(.)$ is a non-decreasing
  function. We  also show below that $C(.)$ is concave. These  facts will be used in proving that \eqref{impe} is the capacity of the system.

To show concavity, for $s_1, s_2 >0$ and $0 \le \lambda \le 1$ we want to show that $C( \lambda s_1 + (1-\lambda) s_2) \ge \lambda C(s_1)+(1-\lambda) C(s_2)$. For $s_i$, let $C_i$ be the capacity achieving codebook, $i=1,2$. Use $\lambda$  fraction of time $C_1$ and $1-\lambda$ fraction $C_2$. Then the rate achieved is $\lambda C(s_1)+(1-\lambda) C(s_2)$ while the average energy used is $\lambda s_1+(1-\lambda) s_2$. Thus, we obtain the inequality showing concavity.

{\bf Theorem 2} For the energy harvesting system with processing energy,

\begin{equation}
C= \sup_{p_x: E[b(X)] \le E[Y]} I(X;W) \label{impee}
\end{equation}
is the capacity for the system.

{\bf Proof:}  :  We prove this result in three steps. First we show the converse i.e., that the achievable rate cannot exceed \eqref{impee}. Next we show that this rate can be achieved by an $iid$ signaling. Finally we show that we can achieve this rate by a signaling scheme that satisfies the energy constraints.

The converse follows via Fano's inequality as in, for example, \cite{cover}, for an AWGN channel. For that proof to hold here, we need that $C(.)$ is concave. 

Next we show that an $iid$ input sequence $\{X_k'\}$ satisfying $E[b(X)]=E[Y]$ achieves the capacity. This can be shown in the usual way by using the following coding-decoding scheme.

{\it{Coding}}: Generate a code book of size $n$ with codewords obtained from an $iid$ sample with the capacity achieving distribution $p_x$  with constraint $E[b(X)]=E[Y]-\epsilon$, where $ \epsilon>0$ is a small constant.

To transmit message $m$, take the corresponding codeword from the above codebook. If the codeword is $\epsilon-$weakly typical and $\sum_{i=1}^n b(x_i)/n \le E[Y]-\epsilon$, then transmit it; otherwise send an error message.

{\it{Decoding}} : On receiving $W^n$, the decoder finds the message $m'$ which has its codeword jointly $\epsilon-$weakly typical with $W^n$, if there is a unique such message. Otherwise it declares an error.

By the usual methods with the above coding-decoding scheme and also  the fact that $C(.)$ is non-decreasing, we can show that the probability of error for this scheme goes to zero as $n \to \infty$. Thus we can achieve the capacity \eqref{impee}.

Finally we show that we can achieve \eqref{impee} by a signaling scheme satisfying the energy constraints. Take a codebook generated by $iid$ sequences given above. Now define
\begin{eqnarray*}
X_k= \begin{cases}
\min\{X_k', \sqrt{(E_k-Z_k)^{+}}\},~& \text{if $X_k'\ge 0$},\\
\max\{X_k', -\sqrt{(E_k-Z_k)^{+}}\},~& \text{if $X_k'< 0$}.
\end{cases}
\end{eqnarray*}
Then to transmit $X_k$ we need energy $T_k= (X_k^2+Z_k)1_{\{X_k \ne 0\}}$ and $E_{k+1}= (E_k-T_k)+Y_k$. Also,
\begin{eqnarray*}
E[T_k]&=& E[X_k^2]+E[Z_k]P\{X_k \ne 0\}\\
&\le& E[X_k'^2]+\alpha P\{X_k' \ne 0\}\\
&=& {E[Y]}-\epsilon.
\end{eqnarray*}

Therefore, as in Theorem 1, $E_k \to \infty~a.s.$ and hence $E_k-Z_k \to \infty~a.s.$ and $|X_k-X_k'| \to 0~a.s.$ Thus $\{X_k\}$ is AMS with the limiting process the $iid$ sequence $\{X_k'\}$. Thus, as before, $\overline{I}(X;W) \ge I^*(X;W)=I^*(X';W')=\overline{I}(X';W')$.
Therefore, we can achieve capacity in \eqref{impee} by the input sequence $\{X_k\}$.~~~~~~~~~~~~~~~~~~~~~~~~~~~~~~~~~~~~~~~~~~~~~~~~~~~~~~~~~~~~~{\raggedleft{$\blacksquare$}}

\vspace{0.1cm}


It is interesting to compute the capacity \eqref{impee} and the capacity achieving distribution $p_X$. Without loss of generality, we can say that under $p_X$, with probability $p ~(0 \le p \le 1)$ the node sleeps and with probability $1-p$ it transmits with a density $f$ that satisfies the cost function with equality. We can show using Kuhn-Tucker conditions that density $f$ is
\begin{equation*}
f(a)=\left(k_1e^{-a^2k_2}-\frac{p e^{-a^2/2\sigma^2}}{(1-p)\sqrt{2\pi\sigma^2}}\right)^{+}
\end{equation*}
where $k_1$ and $k_2$ are positive constants such that $f$ satisfies the cost. When $E[Y]$ is large compared to $E[Z]$ $f$ is close to $\mathcal{N}(0,E[Y]-E[Z])$.

 Then density $f$ can be computed numerically and in general it is not Gaussian.

To get further insight, let $\{B_k\}$ be $iid$ binary random variables with $P[B_1=0]=p=1-P[B_1=1]$ and let $\{G_k\}$ be $iid$ random variables with density $f$. Then $X_k'=B_kG_k$ is the capacity achieving $iid$ sequence. Also,

\begin{eqnarray*}
I(X_k';X_k'+N_k)= h(B_kG_k+N_k)-h(N_k)~~~~~~~~~~~\\
= h(B_kG_k+N_k)-h(B_kG_k+N_k|B_k)~~~~~~~~~~~\\
+h(B_kG_k+N_k|B_k)-h(N_k)\\
= I(B_k;B_kG_k+N_k)+I(G_k;B_kG_k+N_k|B_k)~~\\
=I(B_k;B_kG_k+N_k)+ (1-p)I(G_k;G_k+N_k).~
\end{eqnarray*}
This representation suggests the following interpretation (and coding theoretic implementation) of the scheme: the overall code is a superposition of a binary ON-OFF code and an $iid$  code with density $f$. The position of the ON (and OFF) symbols is used to reliably encode $I(B;BG+N)$ bits of information per channel use, while the code with density $f$ (which is used only during the ON symbols) reliably encodes $(1-p)I(G;G+N)$ bits of information per channel use.

\begin{figure}[ht]
\centering
\includegraphics [height=2.5 in, width=3.5in ]{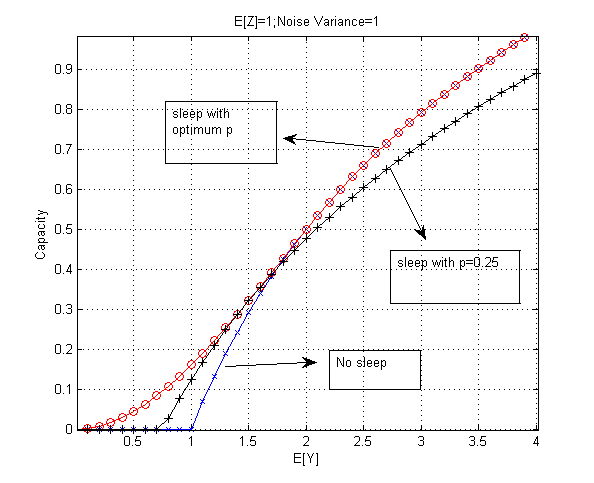}
\caption{Comparison of Sleep Wake policies}
\label{figsleep}
\end{figure}

In Fig.\ref{figsleep} we compare the sleep-wake policies for  $p=0$ (equation \eqref{abc}) and $p= 0.25$ with the optimal $p$. We take $E[Z]=1$ and $\sigma^2=1$.

We see that when $E[Y]$ is comparable or less than $E[Z]$ then the node chooses to sleep with a high probability. When $E[Y]>>E[Z]$ then the node will not sleep at all. Also it is found that when $E[Y]<E[Z]$, the capacity is zero when the node does not have a sleep mode. However we obtain a positive capacity if it is allowed  to sleep.

\section{Achievable Rate with Energy Inefficiencies}
\label{opt}
In this section we make our model more realistic by taking into account the
inefficiency in storing energy in the energy buffer  and the leakage from the energy buffer (\cite{jiang})
 for HSU architecture. For simplicity, we will ignore the energy $Z_k$ used for sensing and processing.

We assume that if energy $Y_k$ is harvested at time $k$, then only energy $\beta_1Y_k$ is stored in the buffer
  and  energy $\beta_2$ gets leaked in each slot where $0<\beta_1 \le 1$ and $0<\beta_2<\infty$.
 Then \eqref{eqn2} becomes
\begin{eqnarray}
E_{k+1}  = ((E_k - T_k)-\beta_2)^{+} + \beta_1Y_k. \label{eqne1}
\end{eqnarray}

The energy can be stored in a super capacitor and/or in a battery. For a supercapacitor,
$\beta_1 \ge 0.95$ and for the Ni-MH battery (the most commonly used battery) $\beta_1 \sim 0.7$.
 The leakage  $\beta_2$ for the super-capacitor as well as the battery is close to
 0 but for the super capacitor it may be somewhat larger.

In this case, similar to the achievability of Theorem 1 we can show that

 \begin{eqnarray}
 R_{HSU} = 0.5~\log\left(1+\frac{\beta_1 E[Y]-\beta_2}{\sigma^2}\right) \label{eqnr1}
 \end{eqnarray}
 is achievable.
This policy is neither capacity achieving nor throughput optimal \cite{vinod1}.
An achievable rate of course is \eqref{fin}. Now one does not even store energy and $\beta_1,~ \beta_2$ are not effective. The upper bound  $0.5~\log(1+{E[Y]} / {\sigma^2})$ is achievable if  $Y$ is chi-square distributed with degree 1.

 In  \emph{Harvest-Use} architecture since the energy harvested is used immediately, there is no loss due to storage inefficiency and leakage. Thus the achievable rate does not depend on $\beta_1$,   $\beta_2$. Hence,  unlike Section III, the rate achieved by the HU may be larger than \eqref{eqnr1} for certain range of parameter values and distributions.

 Another achievable policy for the system with  an energy buffer   with storage inefficiencies is to use the harvested energy $Y_k$ immediately instead of storing in the buffer. The remaining energy after transmission is stored in the buffer. We call this \emph{Harvest-Use-Store} (HUS) architecture. For this case, \eqref{eqne1} becomes
  \begin{eqnarray}
E_{k+1}  = ((E_k + \beta_1(Y_k-T_k)^{+}- (T_k-Y_k)^{+})^{+}-\beta_2)^+. \label{eqne2}
\end{eqnarray}
Compute the largest constant $c$ such that $ \beta_1 E[(Y_k-c)^{+}] >  E[(c-Y_k)^{+}]+
 \beta_2$. This is the largest $c$ such that taking $E[T_k] \le c-\epsilon$
will make $E_k \to \infty~a.s.$ 
Thus, as in
Theorem 1, we can show that rate
 \begin{eqnarray}
 R_{HUS} = 0.5~\log\left(1+\frac{c}{\sigma^2}\right) \label{eqnc1}
 \end{eqnarray}
 is achievable for this system. This is achievable by an input with distribution $iid$ Gaussian with mean zero and variance $c$.

Equation \eqref{eqne1} approximates the system where we have only rechargable battery while \eqref{eqne2} approximates the system where the harvested energy is first stored in a supercapacitor and after initial use transferred to the battery.

When  $\beta_1=1, \beta_2=0$ we have obtained the capacity of this system in Section III. For the general case, its capacity is an open problem although we conjecture that \eqref{eqnc1} may be the capacity.

We illustrate the achievable rates mentioned above via an example.

\subsection{Example 1}

\begin{figure}[!h]
\centering
\includegraphics [height=2.5 in, width=3.5in ]{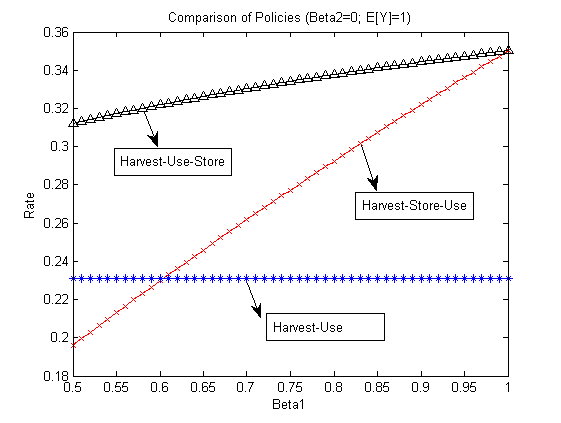}
\caption{Rates for various architectures}
\label{figineff}
\end{figure}

Let $\{Y_k\}$   be $iid$ taking values in  $\{0.25,0.5,0.75,1\}$ with equal probability. We take the
 loss due to leakage $\beta_2=0$. In Figure \ref{figineff} we compare the various architectures discussed in this section for
varying storage efficiency $\beta_1$. We use the result in \cite{peak3} for computing the capacity in \eqref{fin}.

From the figure it can be seen that if the storage efficiency is very poor it is better to use the $ HU$ policy. This requires
no storage buffer and has a simpler architecture. If the
storage efficiency is good $HUS$ policy gives the best performance.
For $\beta=1$, the $HUS$ policy and
$HSU$ policy have the same performance. Unlike the ideal system, the $HSU$ (which uses infinite energy buffer) performs worse than the $HU$  (which uses no energy buffer) when storage efficiency is poor.

Thus if we judiciously use a combination of a super capacitor and a battery, we may obtain a better performance.

\section{Conclusions}
\label{conclude}
In this paper  the Shannon capacity of  an energy harvesting sensor node transmitting over an AWGN Channel is provided. It is shown that the  capacity achieving policies are related to the throughput optimal policies. Also, the capacity is provided when energy is consumed in activities other than transmission. Achievable rates are  provided when there are inefficiencies in energy storage.
\bibliographystyle{IEEEtran}

\bibliography{mybibfilefade}
\end{document}